\documentclass[graybox,natbib,numbers]{svmult}

\usepackage{type1cm}        % activate if the above 3 fonts are
\usepackage{makeidx}         % allows index generation
\usepackage{graphicx}        % standard LaTeX graphics tool
\usepackage{graphicx}	% Including figure files and for \animategraphics
    \graphicspath{{Figures/}}
\usepackage{multicol}        % used for the two-column index
\usepackage[bottom]{footmisc}% places footnotes at page bottom

\usepackage{newtxtext}       % 
\usepackage[varvw]{newtxmath}       % selects Times Roman as basic font

\usepackage[normalem]{ulem} % for \markoverwith, \ULon
\usepackage[version=4]{mhchem}

\usepackage{pdflscape}
\usepackage{afterpage}

\usepackage[T1]{fontenc}

\usepackage{hyperref} 
	\hypersetup{colorlinks,citecolor=blue,linkcolor=blue,urlcolor=blue}
	\usepackage{etoolbox}
		\makeatother

\usepackage[capitalise]{cleveref}
	\crefname{equation}{Equation}{Equations}
	\crefname{figure}{Figure}{Figures}
	\crefname{table}{Table}{Tables}

\newcommand{\pc}{\; {\mathrm{pc}}}				% pc
\newcommand{\cm}{\; {\mathrm{cm}}}				% cm
\newcommand{\mm}{\; {\mathrm{mm}}}				% mm
\newcommand{\mum}{\; \mu {\mathrm{m}}}			% micrometer
\newcommand{\nm}{\; {\mathrm{nm}}}				% nanometer
\newcommand{\K}{\; {\mathrm{K}}}					% K
\newcommand{\au}{\; {\mathrm{au}}}					% au
\newcommand{\yr}{\; {\mathrm{yr}}}					% year
\newcommand{\Myr}{\; {\mathrm{Myr}}}				% Myr
\setcitestyle{notesep={; }}

\DeclareRobustCommand\rsout{%
  \bgroup
  \markoverwith{\textcolor{red}{\rule[0.5ex]{2pt}{0.4pt}}}%
  \ULon
}

\makeindex             % used for the subject index
\begin{document}

\title*{Dust Processing in Protoplanetary Discs From Infall to Dispersal: the Origin of Solar System Isotopic Heterogeneities}
\titlerunning{Dust Processing in Protoplanetary Discs}
\author{
    Mark A. Hutchison\orcidID{0000-0003-4543-8711} and\\
    Maria Sch{\"o}nb{\"a}chler\orcidID{0000-0003-4304-214X} and\\  
    Lucio Mayer\orcidID{0000-0002-7078-2074} and\\ 
    Jean-David Bod\'{e}nan\orcidID{0000-0002-7580-6311}}
\authorrunning{Hutchison et al.}
\institute{Mark A. Hutchison \at Department for Technical Systems, Processes and Communication, Munich University of Applied Sciences HM, Lothstra{\ss}e 34, 80335 M{\"u}nchen, Germany, \email{markahutch@gmail.com}
\and Maria Sch{\"o}nb{\"a}chler \at Institut f\"{u}r Geochemie und Petrologie, Eidgen\"{o}ssische Technische Hochschule Z{\"u}rich, Claussiustrasse 25, CH-8092 Z{\"u}rich, Switzerland, \email{mariasc@ethz.ch}
\and Lucio Mayer \at Institute for Computational Science, University of Z{\"u}rich, Winterthurerstrasse 190, CH-8057 Z{\"u}rich, Switzerland, \email{lucio.mayer@uzh.ch}
\and Jean-David Bod\'{e}nan \at European Science Foundation, 1, quai Lezay-Marn{\'e}sia, 67080 Strasbourg Cedex, France, \email{jdbodenan@esf.org}}
\maketitle

% \abstract*{Each chapter should be preceded by an abstract (no more than 200 words) that summarizes the content. The abstract will appear \textit{online} at \url{www.SpringerLink.com} and be available with unrestricted access. This allows unregistered users to read the abstract as a teaser for the complete chapter.
% Please use the 'starred' version of the \texttt{abstract} command for typesetting the text of the online abstracts (cf. source file of this chapter template \texttt{abstract}) and include them with the source files of your manuscript. Use the plain \texttt{abstract} command if the abstract is also to appear in the printed version of the book.}

\abstract{
The nucleosynthetic heterogeneity between different asteroids and planets is well established. These isotopic variations manifest themselves at the part per millions level or larger, in isotopes that were synthesised in various stellar environments. To escape homogenisation, some of these isotopic signatures must have been preserved in dust, which ended up being heterogeneously distributed in the solar protoplanetary disc.
The origin of the nucleosynthetic heterogeneity is still poorly constrained, potentially reflecting inherited isotope variations from the Sun’s parental molecular cloud and/or processing and redistribution during the subsequent protoplanetary disc phase with thermal processing and size sorting as major processes.
This chapter aims to provide a broad review of the dynamical, collisional, and thermal processes in protoplanetary discs -- from initial infall to gas dispersal -- that may have influenced the distribution and survival of the anomalous carrier phases, which finally accreted into asteroids and planets. While several of these mechanisms have been considered in past studies, they are often examined in isolation, which impedes the assessment of how their effects may be altered or amplified by additional disc processes. Size sorting in particular has received little attention, and here we highlight that this process likely occurred in the disc and can induce nucleosynthetic heterogeneity.
By placing previous studies within the context of a comprehensive overview, we aim to clarify the broader physical framework in which anomalous carrier transport occurs and identify previously underexplored mechanisms that may have contributed to the final isotopic structure of the Solar System we see today.
}

%%%%%%%%%%%%%%%%%%%%%%%%%%%%%%%%%%%%%%%%%%%%%%%%%%

%%%%%%%%%%%%%%%%% BODY OF PAPER %%%%%%%%%%%%%%%%%%

%%%%%%%%%%%%%%%%
\section{Introduction}
\label{sec:introduction}
%%%%%%%%%%%%%%%%

The nucleosynthetic isotope composition of each planetary body is largely unique, serving as a distinct fingerprint for planets and asteroids -- with rare exceptions, such as the Earth-Moon system. These isotopic signatures are powerful tools for tracing genetic relationships between meteorites, extraterrestrial samples returned by space mission, and planets such as Earth and Mars \citep[e.g.][]{Yokoyama/etal/2023,Barnes/etal/2025}. They also offer insights into the source regions, where rocky bodies accreted \citep[e.g.][]{Rufenacht/etal/2023} as well as transport and mixing processes in the solar protoplanetary disc. Together with other cosmochemical data, nucleosynthetic compositions help constrain the mechanisms of planet formation. For example, this approach has been used to determine the lifespan of major reservoirs in the disc \citep{Kruijer/etal/2017}, with implications for the formation and accretion history of Jupiter \citep[e.g.][]{Kruijer/etal/2017,Alibert/etal/2018,Brasser/Mojzsis/2020}. Integrating cosmochemical evidence with astrophysical models has also advanced understanding of how disc evolution shaped the timing and mechanism of planet formation \citep[e.g.][]{Alibert/etal/2018,Lichtenberg/etal/2021,Johansen/etal/2023}.

\begin{figure}
    \centering{\includegraphics[width=\columnwidth]{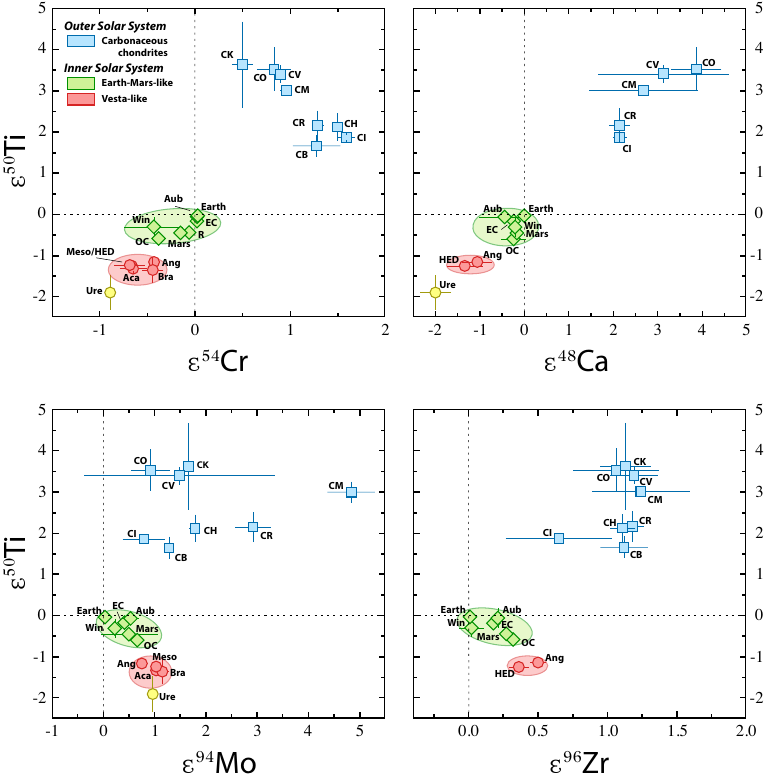}}
    \caption{Averages of various meteorite groups for $\varepsilon^{50}$Ti versus $\varepsilon^{48}$Ca, $\varepsilon^{54}$Cr, $\varepsilon^{94}$Mo and $\varepsilon^{96}$Zr. Quantities are given in $\varepsilon$ notation, defined as the deviation (in parts per 10\,000) of a sample isotope ratio from the corresponding terrestrial standard ratio. Data of supernovae-derived variations ($\varepsilon^{48}$Ca, $\varepsilon^{54}$Cr, $\varepsilon^{50}$Ti) correlate with each other. In contrast, \textit{s}-process variations ($\varepsilon^{94}$Mo, $\varepsilon^{96}$Zr) show a negative correlation with $\varepsilon^{50}$Ti in the inner Solar System and more scatter for carbonaceous chondrites (CC). In addition to the dichotomy between non-carbonaceous (NC) (green, red, yellow symbols) and CC meteorites (blue symbols), the NC reservoir displays three distinct clusters (green, red and yellow) defined by the Earth-Mars like meteorites, the Vesta-like meteorites and ureilites. These have been interpreted as the result of planetesimal formation in different rings in the protoplanetary disc \citep{Rufenacht/etal/2023}. Uncertainties are student-t 95\% confidence intervals. Abbreviations: 
    Aca -- acapulcoites, 
    Ang -- angrites, 
    Aub -- aubrites, 
    Bra -- brachinites,
    (CB, CH, CI, CK, CM, CO, CR, CV) -- classes of CCs,
    EC -- enstatite chondrites, 
    HED -- howardites-eucrites-diogenites
    OC -- ordinary chondrites, 
    Meso -- mesosiderites,  
    R -- rumuruti chondrites,
    Ure -- Ureilites, 
    Win -- winonaites. 
    Figure adapted from \citet{Rufenacht/etal/2023}.}
    \label{fig:rufenacht}
\end{figure}

A primary manifestation of the nucleosynthetic isotope heterogeneity is the isotopic dichotomy -- seen, for example, in Ti and Cr isotopes (\cref{fig:rufenacht}) -- that separates non-carbonaceous (NC) chondrites and related achondrites from the inner Solar System from carbonaceous (CC) chondrites and related achondrites that formed further out \citep{Trinquier/Birck/Allegre/2007,Leya/etal/2008,Trinquier/etal/2009}. This pattern, traceable today in the isotopic compositions of bulk meteorites originating from asteroids and planets, ultimately reflects the distribution of presolar grains (formed around earlier generations of stars), and other isotopically distinct carriers identified in meteorites. Heterogeneities are observed in isotopes of refractory elements \citep[e.g. Ca, Ti, Zr, Mo, Ru, Pd, Nd;][]{Toth/etal/2020,Ek/etal/2020}, as well as some slightly more volatile elements \citep[Ni, Fe, Cr and Zn;][]{Steele/etal/2011,Cook/Schonbachler/2017,Schiller/Bizzarro/Siebert/2020,Rufenacht/etal/2023,Martins/etal/2023}. This indicates that most nucleosynthetic signatures of more volatile elements were homogenised, due to the lack of signature preservation in a carrier, be it that the isotopes were never incorporated into a solid phase after nucleosynthesis, by destruction of labile carriers in the interstellar medium (ISM) or during the protosolar disc formation \citep{Toth/etal/2020,Ek/etal/2020}. For the assessment of the nucleosynthetic data, it is also important to note that processes on the meteorite parent body (e.g. aqueous alteration or heating) can modify these signatures depending on the element and its presolar carrier \citep[e.g., for Cr, Os;][]{Yokoyama/Alexander/Walker/2011,Yokoyama/etal/2023}.

At present, however, it remains uncertain whether this isotopic structure was inherited during infall, generated by processes within the protoplanetary disc phase, or reflects a combination of both. Past studies have identified possible contributions from both mechanisms, including: inherited heterogeneity from the Sun's nascent molecular cloud \citep{Clayton/1982,Dauphas/Marty/Reisberg/2002a,Dauphas/etal/2004,Nanne/etal/2019,Ek/etal/2020}; fractionation of material during infall \citep{Van-Kooten/etal/2016}; subsequent evolution within the circumsolar disc due to thermal processing \citep{Trinquier/etal/2009,Burkhardt/etal/2012,Paton/Schiller/Bizzarro/2013,Akram/etal/2015,Poole/Rehkaemper/Coles/2017,Ek/etal/2020,Colmenares/etal/2024}; dynamical sorting resulting from grain size evolution, aerodynamic drag, and/or viscous evolution \citep{Pignatale/etal/2017,Pignatale/etal/2019b,Hutchison/etal/2022}, and interactions with early Jupiter and the gap it created \citep{Kruijer/etal/2017,Alibert/etal/2018,Homma/etal/2024}. Disentangling these possibilities is a complex task, hampered by the fact that most models are limited in scope. Due to computational constraints and physical uncertainties, simulations often focus on individual processes in isolation rather than attempting to evolve the full, chemically diverse grain population in a realistic disc setting. Thus, even when spatial isotopic variations are produced in a given model, their survival remains uncertain once more complex physics or longer-term evolution through the protoplanetary and debris disc phases are considered. This modelling gap remains a key barrier to understanding the origins of the isotopic variations in our Solar System.

Fortunately, this situation is beginning to change. Besides a growing nucleosynthetic data set, advances in numerical methods and computational capabilities are enabling models to capture increasingly realistic treatments of dust and chemical evolution. For example, \citet{Benitez-Llambay/Krapp/Pessah/2019} introduced a multi-species dust dynamics framework that avoids restrictive assumptions about dust-gas coupling. \citet{David-Cleris/Laibe/Lapeyre/2025} developed a highly parallel GPU-based code optimized for exascale architectures, paving the way for global disc simulations at unprecedented resolution and scale -- essential for capturing instabilities like the streaming instability. Meanwhile, \citet{Lombart/Guillaume/2021} and \citet{Lombart/etal/2024} designed growth and fragmentation solvers that preserve accuracy with relatively few dust bins, enabling models of grain size evolution (and potentially porosity and composition) in three-dimensional hydrodynamic settings. A growing number of models now also track the compositional evolution of dust grains as they migrate through the disc \citep[e.g.][]{Van-Clepper/etal/2022,Eistrup/Cleeves/Krijt/2022,Vaikundaraman/etal/2025}. Such progress is essential if we hope to understand how disc processes may have influenced the distribution of presolar grains in micron-sized dust to kilometre-scale planetesimals.

Without a definitive answer to the origin of isotopic heterogeneity, we must remain open to the possibility that both inherited traits and protoplanetary disc evolution contributed to shaping the isotopic architecture of the Solar System. This section does not aim to identify a single dominant mechanism, but rather to survey the key physical processes within discs that govern the transport, concentration, and transformation of dust -- highlighting their characteristic timescales, the regions of the disc where they are most influential, and their potential isotopic relevance (see \cref{fig:overview_macroscopic}). In doing so, we give particular attention to dust size sorting, which has received comparatively little emphasis in past discussions but likely played a central role in establishing nucleosynthetic variations. We begin with brief overviews of constraints from meteorites, followed by three modes of dust evolution: dynamical, collisional, and thermal. We then turn to a more integrative discussion of how these processes, individually and in concert, may have influenced the isotopic composition of solids within the disc.

The NCCR PlanetS has contributed in several aspects to the characterisation of the nucleosynthetic heterogeneity, its origin and implications. This includes contributions of novel high precision isotope data and their interpretation in the cosmochemical context [e.g. \citealt{Cook/Schonbachler/2017,Cook/etal/2018,Cook/Leya/Schonbachler/2020,Cook/Meyer/Schonbachler/2021,Ek/etal/2020,Rufenacht/etal/2023,Anand/etal/2024,Palme/Mezger/2024,Anand/Mezger/2025,Schonbachler/etal/2025}], as well as their combination with various astrophysical models [e.g. \citealt{Alibert/etal/2018,Lichtenberg/etal/2021,Bodenan/etal/2020,Hutchison/etal/2022,Hunt/etal/2022}] to constrain the evolution of dust in the disc and formation of asteroids and planets, both terrestrial and giant.

\afterpage{%
\clearpage % ensure the landscape page starts cleanly
\begin{landscape}
\begin{figure}[p]
    \centering
    \includegraphics[width=\linewidth,height=\textheight,keepaspectratio]{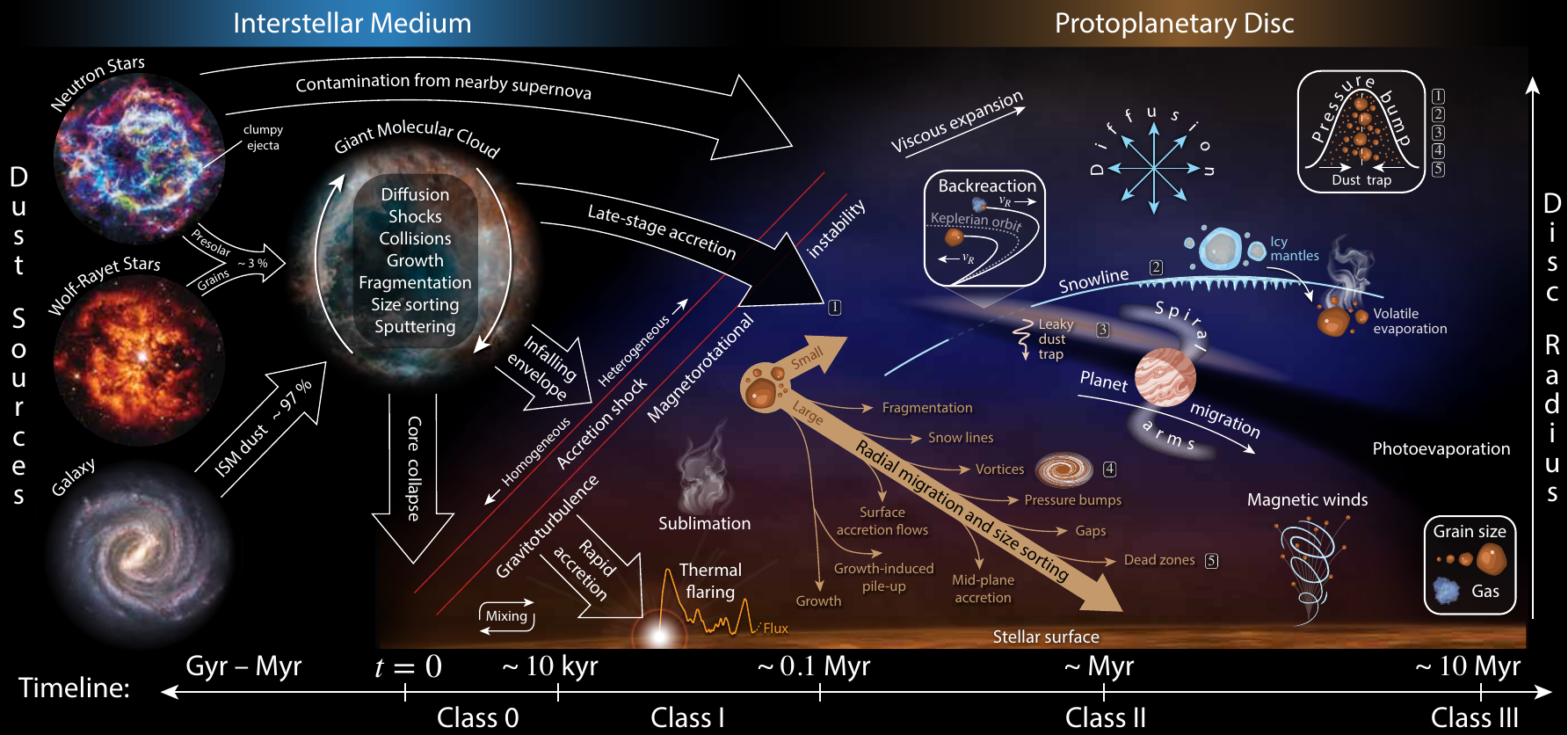}
    \caption{A schematic overview of key processes affecting dust on global scales, from its initial production in the interstellar medium (left) to the end of the protoplanetary disc phase (right). Arrows indicate transport and temporal evolution (horizontal axis), while the vertical axis in the disc representation corresponds to the cylindrical distance from the central star (bottom) to the outer disc (top). Infalling material and gravitoturbulence initially drive strong mixing and rapid accretion onto the star. As infall shifts outward and gradually subsides, the magnetorotational instability becomes the dominant mechanism governing disc evolution. Dust migrates and is aerodynamically size-sorted, with radial drift slowed or halted at local pressure maxima. The dust traps created by these pressure bumps can facilitate the formation of planetesimals, which are crucial building blocks of planets. Solids undergo varying degrees of thermal processing depending on distance from the star and transient heating events. Collectively, these processes influence isotopic heterogeneity, volatile depletion, and the spatial distribution of solids, ultimately shaping the composition of planetesimals and planetary bodies that emerge from the protoplanetary disc phase.
    }
    \label{fig:overview_macroscopic}
\end{figure}
\end{landscape}
\clearpage % end landscape cleanly
}
%

%%%%%%%%%%%%%%%%
\section{Constraints from meteorites}
\label{sec:constraints_from_meteorites}
%%%%%%%%%%%%%%%%

Some of the most pristine samples of dust and aggregates from the solar protoplanetary disc are preserved only in chondrites, the most primitive meteorites.\footnote{Note cometary dust (i.e. interplanetary dust particles collected in the stratosphere or returned by the STARDUST mission) is likely even more primitive, but the grains are typically too small to allow high-precision bulk isotopic analyses.}
They represent fragments of undifferentiated (never fully molten) planetesimals or the outermost layers of differentiated ones \citep{Scott/Krot/2014,Elkins-Tanton/etal/2011}. In contrast, achondrites derive from differentiated bodies that were heated to temperatures sufficient to melt metal and (at least partially) silicates, leading to the formation of a metallic core and a silicate mantle. This process altered the originally accreted dust, rendering it no longer directly accessible. As a result, differentiated bodies generally exhibit a single nucleosynthetic isotopic composition, since melting destroyed and homogenised the original isotopic carriers of nucleosynthetic variations. Both chondrite and most achondrite parent bodies formed early, within the first few million years of Solar System evolution \citep{Wadhwa/2014}.

\begin{figure}
    \centering{\includegraphics[width=0.5\columnwidth]{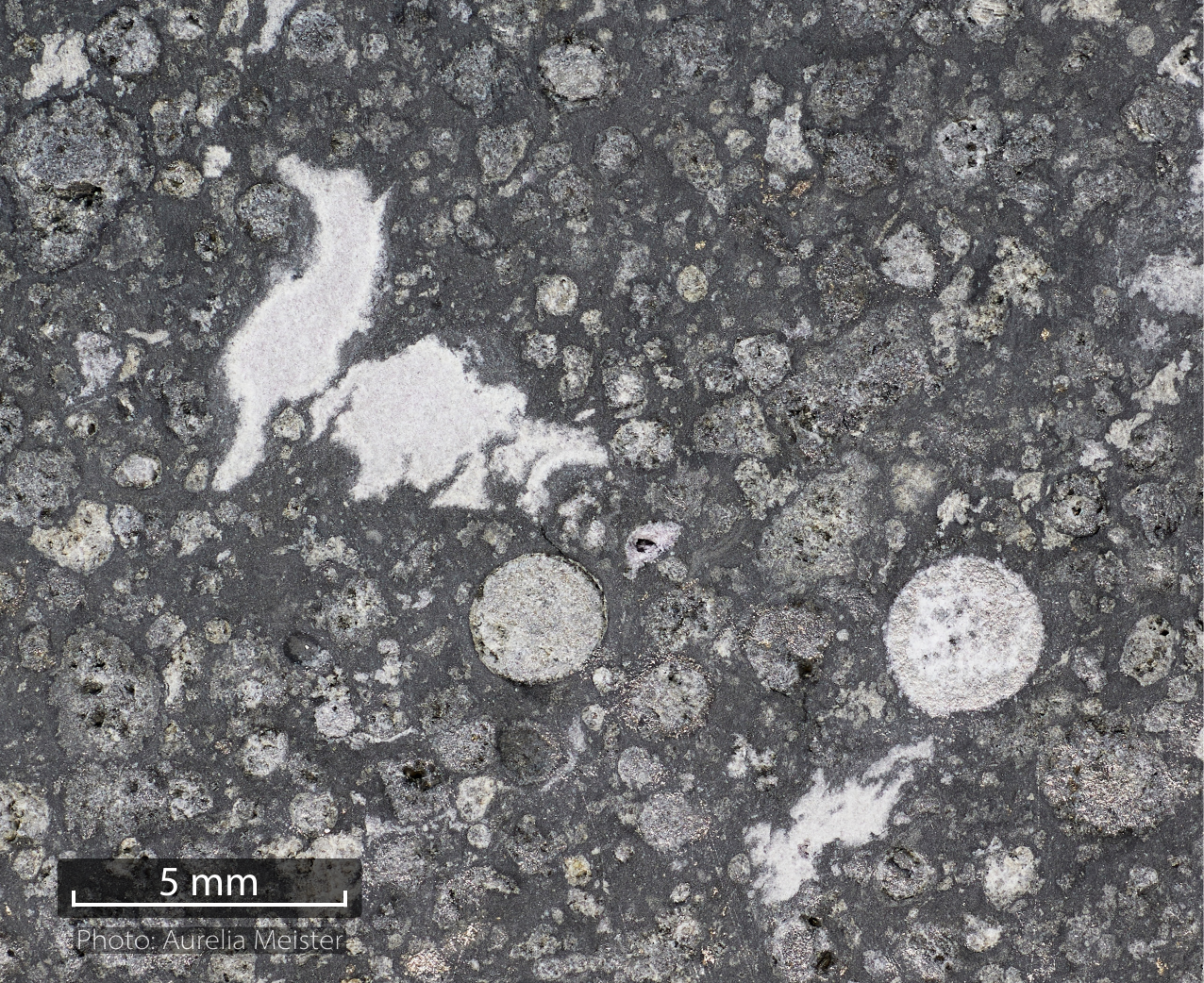}}
    \caption{
    A cross-section through the carbonaceous chondrite Allende prominently showing several white fluffy CAIs and one large white round CAI. Darker, spherical chondrules are also clearly visible. The matrix is the dark fine-grained, interstitial material. Image courtesy of Aur{\'e}lia Meister, ETH Meteorite Collection (used with permission).
    }
    \label{fig:meteorite}
\end{figure}

Since chondrites have largely preserved their originally accreted material, they act as time capsules of the disc’s primordial dust. Nevertheless, depending on the meteorite type, the level of parent-body alteration can still range from minor to extensive \citep{Scott/Krot/2014}. Careful sample selection is therefore essential to recover information about the original dust from the disc. Chondrites contain a wide variety of components (see \cref{fig:meteorite}) and minerals, whose proportions and occurrences vary between chondrite groups. Their principal components are (i) refractory inclusions, such as Calcium-Aluminium-rich Inclusions (CAIs) and Amoeboid Olivine Aggregates (AOAs); (ii) chondrules, once-molten silicate spherules formed in transient high-temperature events and cooled rapidly over hours to days; and (iii) matrix, a mixture of fine-grained material ($<1$--$10\mum$) that includes, among other phases, silicates, metal, sulfides, organics, and presolar grains.

\begin{description}
    \item[\textit{Calcium–aluminium-rich inclusions (CAIs):}]
    The earliest preserved solids of our Solar System \citep[e.g.][]{Amelin/etal/2010,Connelly/etal/2012} -- provide direct evidence for thermal processing and transport in the disc. They have elemental compositions consistent with the first 5\% of solids expected to condense from a gas of solar composition \citep{Davis/Richter/2014}. CAIs represent aggregates of high-temperature condensates that formed near the protosun \citep[e.g.][]{Krot/2019}, and some record prolonged formation histories with multiple heating and melting events in the disc \citep[e.g.][]{Krot/etal/2007,MacPherson/etal/2012,Kita/etal/2013,Larsen/etal/2025}. Their accretion together with organics and ices onto the parent bodies of chondrites implies that CAIs were transported outward from the hot inner regions near the Sun to colder environments beyond the water snow line, where carbonaceous chondrites formed. Once incorporated into these asteroidal parent bodies, subsequent low-temperature aqueous alteration and thermal metamorphism further modified the chemical and isotopic compositions of CAIs to varying degrees \citep{Weisberg/etal/1993,Schrader/etal/2014}.
    
    \item[\textit{Chondrules:}]
    Most chondrules formed $1$--$3\Myr$ after CAIs \citep[e.g.][]{Kita/etal/2005,Schonbachler/etal/2025} during repeated brief, local flash-heating events reaching peak temperatures of $\sim1750$--$2150 \K$, followed by rapid cooling \citep[e.g.][]{Rubin/2000}. Chondrules formed in both the NC and CC regions of the disc, with those in the CC region forming, on average, later \citep{Fukuda/etal/2022}. Mean chondrule diameters range from $0.2$ to $>1\mm$, with each chondrite group exhibiting a characteristic, albeit not mutually exclusive, size range \citep{Jones/2012}. These size distributions were likely shaped by aerodynamic size-sorting in the disc \citep{Cuzzi/etal/2001}. 
    
    \item[\textit{Matrix (including presolar grains):}]
    The fine-grained matrix filling the space between chondrules in chondrites is composed primarily of silicates, including chondrule fragments (e.g., olivine and pyroxene) and amorphous silicates, together with sulfides and additional phases such as Fe–Ni metal and carbonates (depending on meteorite class), as well as minor components including organic compounds and presolar grains. The matrix is often aqueously altered to varying degrees, causing the original minerals to transform into phyllosilicates, magnetite, and other minerals through mild parent body heating, melting of accreted ices, and reactions with the silicates \citep{Lee/etal/2025}.

    \par\hspace*{1em}
    Presolar grains are a minor (<ppm to 100s ppm) component of the matrix that formed around earlier generations of stars and are identified by their highly anomalous isotopic compositions indicative of their stellar formation site. They can be modified and destroyed by parent-body processes \citep[aqueous alteration and thermal metamorphism; e.g.][]{Huss/Lewis/1995}. Organic matter, mostly macromolecular, is even more sensitive to temperature \citep[e.g.][]{Busemann/Alexander/Nittler/2007}. The existence of both presolar grains and organics in chondrites provides evidence for low degrees of thermal processing of the matrix. Hence the volatile-rich matrix stands in stark contrast to refractory chondrules and CAIs, the latter providing direct evidence for high-temperature reprocessing under diverse conditions across different regions of the disc. This contrast is also supported by the depletion trend of moderately volatile elements in carbonaceous chondrites relative to the solar composition, which is mirrored by the amount of thermally processed material (refractory inclusions and chondrules) versus volatile-rich matrix \citep{Braukmuller/etal/2018,Morton/etal/2024}. 
\end{description}

The components of chondrites show a hierarchy in nucleosynthetic isotope variations: the most extreme occur in presolar grains \citep[e.g.][]{Liu/2025}, followed by CAIs, then chondrules exhibiting more subtle variations \citep[e.g.][]{Bermingham/etal/2020}. This pattern indicates that the originally large isotopic heterogeneities carried by presolar grains were progressively reworked and homogenized through (i) high-temperature recycling, such as evaporation and condensation processes close to the Sun (e.g. CAI formation), and (ii) chondrule formation, where mixtures of presolar grains and solar nebular dust were melted and fused into new objects. Moreover, both CAIs and chondrules provide evidence for repeated recycling of earlier generations of dust, including other chondrule and CAI fragments \citep{Krot/2019,Marrocchi/etal/2024}.

Interpreting bulk isotope heterogeneity in meteorites is complicated by the fact that all meteorites derive from parent bodies that experienced varying degrees of aqueous alteration and/or thermal overprinting of the originally accreted material. These processes can directly erase presolar grains. For example, \citet{Huss/Lewis/1995} showed that presolar SiC and graphite were destroyed during increasing thermal metamorphism on chondrite parent bodies. Aqueous alteration can have similar consequences: if fluids preferentially dissolve certain presolar phases and locally redistribute their isotopes, this could result in an isotopically heterogeneous body. Consequently, analysed chondrite samples may not faithfully record the bulk isotope composition of their parent body or its accretion region. The extent to which an element is affected by these processes depends on their chemical susceptibility \citep[e.g. as discussed for Os and Cr isotopes;][]{Yokoyama/Alexander/Walker/2011,Yokoyama/etal/2023}. These effects represent a second-order disturbance superimposed on the primary heterogeneities inherited from the dust of the accretion region. In many cases (e.g. Cr isotopes) this disturbance introduces small-scale heterogeneity, but if larger samples of a chondrite are analysed, they can still be used to identify genetic relationships and accretion regions of first-generation parent bodies \citep{Rufenacht/etal/2023,Yokoyama/etal/2023}.

%

%%%%%%%%%%%%%%%%
\section{Dynamical evolution}
\label{sec:dynamical_evolution}
%%%%%%%%%%%%%%%%

\begin{figure}
    \centering{\includegraphics[width=\columnwidth]{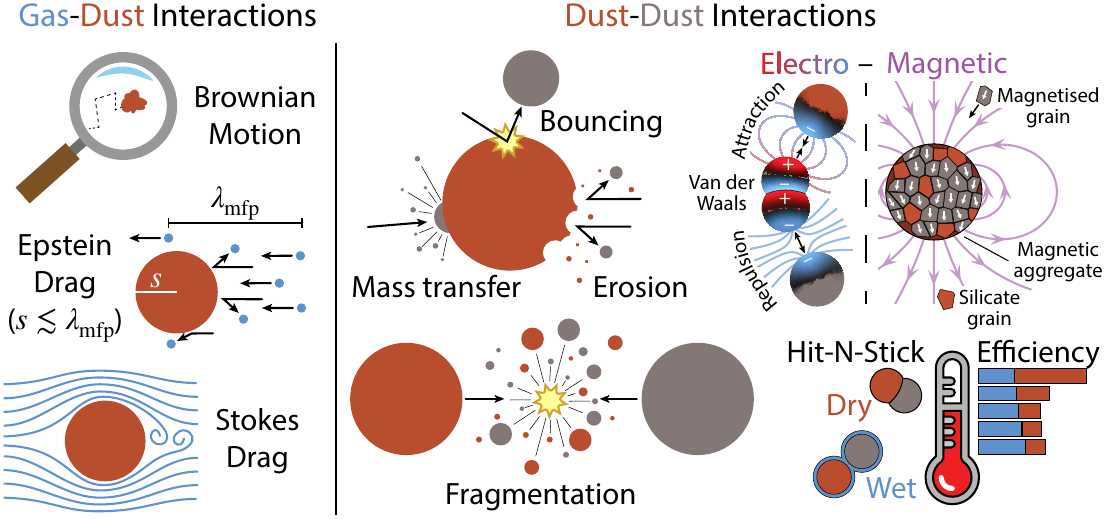}}
    \caption{Illustration of the principal gas–dust and dust–dust interactions that govern the motion and size evolution of individual dust grains in protoplanetary discs. At the smallest scales, Brownian motion sets the random velocities of dust grains, while aerodynamic coupling to the gas occurs through Epstein or Stokes drag, depending on whether the grain size ($s$) is smaller or larger than the gas mean free path ($\lambda_{\rm mfp}$). Differential dust velocities drive migration and size sorting, and determine both the frequency and outcome of collisions, which may result in sticking, bouncing, mass transfer, erosion, or fragmentation. The sticking efficiency depends on surface conditions, temperature, and interparticle forces (van der Waals, electrostatic, and magnetic), which can either aid or hinder aggregate growth. Under wet (ice-coated) conditions, the sticking efficiency decreases with increasing temperature. Dry (ice-free) surfaces are roughly ten times more efficient overall but follow the same declining trend up to $\sim1000\K$ before rising sharply at higher temperatures, as shown schematically by the coloured bars to the right of the thermometer.
    }
    \label{fig:overview_microscopic}
\end{figure}

In the absence of gas and radiation, dust follows Keplerian orbits about the central star. The presence of gas modifies this motion due to drag (see left-hand side of \cref{fig:overview_microscopic}), the magnitude of which is governed by grain size/shape, intrinsic dust density, sound speed and density of the gas, and the differential velocity between phases \citep{Epstein/1924,Baines/Williams/Asebiomo/1965}. These dependencies are encapsulated in the particle stopping time, $t_{\rm s}$, which is short for fine, porous grains in dense gas (e.g. seconds for $\mum$-sized grains in the inner disc), and long for macroscopic, compact bodies in tenuous gas \citep[e.g. decades for $\cm$-size grains in the outer disc or longer than the disc lifetime for planetesimals; see Figure 6 from][]{Price/etal/2018}. The degeneracy caused by the spread of physical values in the above dependencies implies that it is often simpler to use the dimensionless Stokes number, $\text{St} \equiv \Omega_\text{K} t_\text{s}$ (where $\Omega_{\rm K}$ is the Keplerian angular velocity), to consolidate the different dynamical behaviours of dust grains into three important regimes: $\text{St} \ll 1$ (perfectly coupled to the gas), $\text{St} \sim 1$ (partial coupling), $\text{St} \gg 1$ (decoupled from the gas).

%

%%%%%%%%%%%%%%%%
\subsection{Radial migration}
\label{sec:radial_migration}
%%%%%%%%%%%%%%%%

Inward radial migration is a consequence of the headwind dust feels from pressure-supported gas that orbits at sub-Keplerian speeds, causing the dust to lose angular momentum and drift inward \citep{Whipple/1972,Weidenschilling/1977}. This effect is most pronounced for intermediate-sized grains with $\text{St} \sim 1$, and can result in significant loss of $\gtrsim \mm$-sized grains onto the central star \citep{Weidenschilling/1977,Testi/etal/2014}. However, a large host of mechanisms can slow or even halt inward migration of these grains, including grain growth \citep[e.g.][]{Brauer/Dullemond/Henning/2008,Laibe/Gonzalez/Maddison/2008}, transitions from Epstein to Stokes drag \citep[e.g.][]{Birnstiel/etal/2010,Okuzumi/etal/2012}, pile-up by combined effects of gas drag and grain growth \citep[e.g.][]{Stepinski/Valageas/1997,Youdin/Shu/2002}, pile-up at Magneto-Rotational Instability dead zone boundaries \citep{Chatterjee/Tan/2014}, trapping in local pressure maxima \citep{Nakagawa/Sekiya/Hayashi/1986,Pinilla/etal/2012}, azimuthal vortices induced by various fluid instabilities \citep[e.g.][]{Lovelace/etal/1999,Nelson/Gressel/Umurhan/2013,Klahr/Hubbard/2014}, gap formation by planets \citep{Paardekooper/Mellema/2006,Rice/etal/2006}, and turbulent eddies \citep{Gerosa/etal/2024}. As grains grow to sizes that yield Stokes numbers much greater than unity, their radial migration rate declines sharply due to their weak coupling with the gas. They continue to experience a headwind from the sub-Keplerian gas, but they are too massive to be significantly affected by the amount of momentum extracted by the drag.

\begin{figure}
    \centering{\includegraphics[width=0.835\columnwidth]{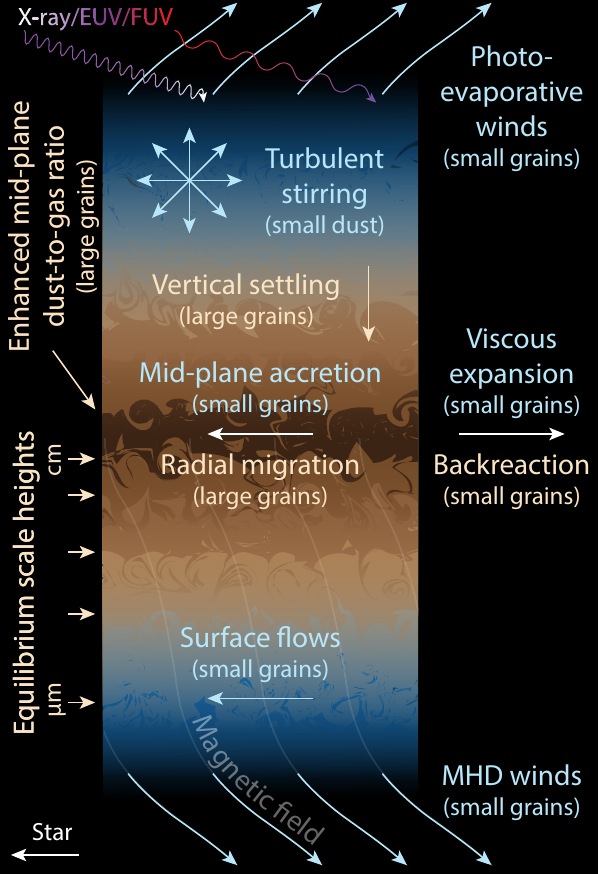}}
    \caption{Schematic vertical cross-section through a localised disc region showing different transport processes for the gas (blues) and dust (browns). Parenthetical notes indicate whether transport processes are more likely to influence small or large grains. For example, large grains experience vertical settling toward the mid-plane and radial migration towards the star. Their finite vertical extents are determined by how strongly they are coupled to the turbulent gas motions. On the other hand, small grains are well-coupled to the gas and susceptible to turbulent stirring, viscous expansion, backreaction of the gas from larger dust grains, and entrainment in disc winds. Photoevaporative winds are thermally driven flows launched when high-energy radiation heats the disc atmosphere. In contrast, magnetohydrodynamic (MHD) winds are driven by magnetic fields threading the disc, which eject ionised gas near the disc surface along field lines. Both types of winds can entrain small dust grains, removing them or redistributing them to larger radii. However, MHD winds additionally remove angular momentum from the disc, producing inward surface flows that transport small grains into the inner disc.
    }
    \label{fig:vertical_cross-section}
\end{figure}

In contrast, the tightly coupled grains with small Stokes numbers tend to trace the viscous evolution of the gas, being carried both inward by accretion flows and outward by viscous expansion \citep[often with a net outward migration rate; see][]{Dipierro/etal/2018}. In regions with high dust-to-gas ratios, the outward movement can be further enhanced by the backreaction of the dust on the gas \citep{Bai/Stone/2010a,Hutchison/etal/2022}. More generally, observational evidence also supports outward radial transport of solids. Crystalline silicates -- which require formation temperatures $\gtrsim 1000 \K$ and are thought to form in the inner disc -- are detected well beyond their expected formation region \citep{Keller/Gail/2004, Ciesla/2009}, with transient features that change on timescales of months to decades \citep{Abraham/etal/2009}. Moreover, isotopic and petrographic studies of chondrules from certain meteorites (carbonaceous chondrites) show that they contain an assortment of both inner and outer Solar System material, implying that large-scale outward transport of $\mm$-sized solids occurred despite radial barriers such as Jupiter \citep{Williams/etal/2020}. Outward advection and turbulent diffusion \citep[e.g.][]{Hughes/Armitage/2010}, magnetic pressure-driven jets \citep[e.g.][]{Liffman/etal/2020}, and disc winds \citep[e.g.][]{Owen/Ercolano/Clarke/2011a} have all been proposed as potential transport mechanisms. The radial motions of gas and dust, as well the vertical motions discussed in the next section, are illustrated together in \cref{fig:vertical_cross-section}.

%%%%%%%%%%%%%%%%
\subsection{Vertical settling}
\label{sec:vertical_settling}
%%%%%%%%%%%%%%%%

The lack of pressure support in the dust also influences its vertical motion. While the gas maintains vertical structure through hydrostatic balance with stellar gravity, the unbalanced vertical gravitational force on the dust, combined with drag from its differential motion relative to the gas, causes it to settle toward the mid-plane like an (over-)damped harmonic oscillator \citep[e.g.][]{Nakagawa/Sekiya/Hayashi/1986}. Because settling timescales are typically much shorter than those for radial drift, the two processes can often be treated independently \citep{Laibe/Gonzalez/Maddison/Crespe/2014}. Sedimentation increases the dust density near the mid-plane, enhancing grain growth \citep{Youdin/Lithwick/2007} and modifying the radial migration behaviour of solids. Under certain conditions, the elevated dust-to-gas ratio can trigger collective instabilities, including the streaming instability \citep{Youdin/Goodman/2005} or gravitational instability \citep{Youdin/Shu/2002,Youdin/2011}, which can lead to the formation of planetesimals \citep[e.g.][]{Drazkowska/Dullemond/2014}.

In turbulent protoplanetary discs, aerodynamic coupling between dust and gas can sustain a finite vertical distribution of solids \citep[e.g.][]{Dubrulle/Morfill/Sterzik/1995}. Although the strength and structure of turbulence remain challenging to constrain observationally, the consequences of dust settling -- such as reduced disc flaring and vertical dust stratification -- are relatively well understood \citep[e.g.][]{Dullemond/Dominik/2004}. These features are generally consistent with observations of spectral energy distributions and resolved disc images \citep[e.g.][]{Dullemond/Dominik/2004,Rilinger/etal/2023}. Turbulent gas motions also induce diffusion of dust through drag forces \citep[e.g.][]{Youdin/Lithwick/2007}, which acts not only to support the vertical structure of the dust layer but also to redistribute grains radially and vertically \citep[e.g.][]{Hutchison/etal/2022}. This is particularly true for small fragments generated by collisional processes in the mid-plane \citep[e.g.][]{Dullemond/Dominik/2005}, which can escape dust traps \citep{Drazkowska/etal/2019} and cross planetary gaps \citep{Stammler/etal/2023}.

%

%%%%%%%%%%%%%%%%
\section{Collisional evolution: grain growth and destruction}
\label{sec:collisional_evolution}
%%%%%%%%%%%%%%%%

Observational evidence for mm- to cm-sized grains in protoplanetary discs strongly supports the idea that solids grow incrementally through collisions (see right-hand side of \cref{fig:overview_microscopic}). Sticking in these early stages is typically attributed to short-range dipole-dipole interactions, such as van der Waals forces or electrostatic dipoles in water ice \citep[e.g.][]{Blum/Wurm/2008}. While grains are typically assumed to be electrically neutral and nonmagnetic under standard disc conditions, several studies have shown that charge build-up, magnetic interactions, or charge separation can enhance or inhibit sticking in small aggregates \citep[e.g.][]{Nuth/etal/1994,Okuzumi/2009}. In turn, grain growth can influence the ionisation balance and magnetic resistivity of the disc, potentially feeding back into its global evolution \citep{Tsukamoto/Machida/Inutsuka/2023}. Many additional factors affect growth rates, including size, composition, internal structure, impact parameter, and collision velocity \citep{Birnstiel/2024}. Compositionally, silicates, metals, ices, and organics dominate \citep{Lewis/1995}, and while water ice and organic materials can increase sticking efficiency \citep[e.g.][]{Gundlach/etal/2011}, lab studies have focused mostly on silicates \citep{Blum/Wurm/2008}. Local conditions such as condensation near snow lines can further aid growth, especially where they modify disc turbulence and generate pressure bumps that trap dust \citep[e.g.][]{Brauer/Henning/Dullemond/2008}. However, there is simply not enough gas-phase material to grow large mantles on every grain \citep{Testi/etal/2014}.

The microphysics of collisions reveals that grain morphology and porosity often play a more decisive role than composition. Threshold sticking velocities are lower for compact grains and higher for irregular or porous aggregates, with velocities decreasing as size increases \citep{Poppe/Blum/Henning/2000a,Kothe/etal/2013}. Aerodynamic drag further complicates growth by introducing differential velocities between grains of varying sizes, due to size-dependent migration and settling \citep{Nakagawa/Sekiya/Hayashi/1986,Brauer/Dullemond/Henning/2008}. Small grains experience Brownian motion \citep{Blum/etal/1996}, while larger grains respond to turbulent gas fluctuations \citep[e.g.][]{Ormel/Cuzzi/2007}. Growth eventually halts due to bouncing \citep{Zsom/etal/2010}, fragmentation \citep{Dominik/etal/2007}, or erosion \citep{Schrapler/Blum/2011}. Non-Gaussian velocity distributions do allow for rare, favourable collisions to transfer mass to larger grains \citep[e.g.][]{Windmark/etal/2012b}. If aided by processes such as reaccretion of fragments \citep{Wurm/Blum/Colwell/2001a}, charge build-up \citep{Poppe/Blum/Henning/2000b}, and high local dust-to-gas ratios in pressure traps \citep{Meheut/etal/2012,Gonzalez/etal/2015a}, it may be possible for certain grains to grow collisionally to planetesimal sizes -- albeit inefficiently \citep[see e.g.][]{Estrada/Cuzzi/Morgan/2016} and potentially over timescales longer than the disc lifetime. The dominant pathway to planetesimal formation is now widely believed to proceed via the streaming instability, a resonant drag-driven instability that enables dust to concentrate and collapse gravitationally under the right conditions \citep[e.g.][]{Youdin/Goodman/2005,Squire/Hopkins/2018}. Strong clumping within the streaming instability typically requires sufficiently large particles and elevated dust-to-gas ratios near the mid-plane \citep[e.g.][]{Simon/etal/2024}.

%%%%%%%%%%%%%%%%
\section{Thermal processing}
\label{sec:thermal_processing}
%%%%%%%%%%%%%%%%

During infall \citep[typically lasting $\sim$0.1–0.5 Myr;][]{Dunham/etal/2014,Kristensen/Dunham/2018}, refractory grains can generally survive passage through the envelope and accretion shock down to $\sim 1\au$, whereas icy mantles already start sublimating at radii of $\sim 2$–$30\au$, depending on their volatility and the thermal structure of the collapsing system \citep{Chick/Cassen/1997}. This early phase of infall and disc assembly is particularly important, as the interplay between dust transport and thermal processing leads to local mixing of materials with diverse thermal histories \citep{Pignatale/etal/2018}. After infall, heating is governed by viscous dissipation (i.e. the conversion of mechanical energy from shear forces between adjacent fluid layers into heat) near the mid-plane and stellar irradiation in the disc's surface layers and outer regions, generating strong radial and vertical temperature gradients \citep{DAlessio/etal/1998,Dullemond/Dominik/2004}. These gradients establish the locations of volatile snow lines, which shift over time as stellar luminosity and accretion rates evolve \citep[e.g.][]{Lichtenberg/etal/2021}. Vertical circulation and turbulent diffusion can lift grains into warmer surface layers, exposing them to elevated temperatures, altered chemical conditions, and photoprocessing \citep{Ciesla/2010}. Conversely, gas-phase volatiles that diffuse back across snow lines (or circulate downward from the surface) can recondense onto dust grains in the mid-plane, enhancing the local ice surface density and promoting planetesimal formation through the so-called cold-finger effect \citep[e.g.][]{Meijerink/etal/2009}. Thermal outbursts from the central star -- such as FU Orionis-type events -- can transiently raise disc temperatures substantially, temporarily shifting snow lines outward and reprocessing solids \citep{Audard/etal/2014,Colmenares/etal/2024}. The rise time of these eruptions is typically on the order of $\sim1$--$10\yr$ or more and can remain in a high-accretion, high-luminosity phase for decades \citep{Herbig/1989,Clarke/etal/2005}. Spiral density waves, triggered by gravitational instabilities or massive planets, can also generate localised heating via shock dissipation, which may exceed background viscous heating and facilitate chondrule formation  \citep{Boley/Durisen/2008,Bodenan/etal/2020,Ono/etal/2025}. This is particularly interesting given that spirals from gravitational instabilities are expected to host overdensities of $\cm$-sized particles \citep{Dipierro/etal/2015a}, placing the raw material for chondrules in the same regions where heating is strongest. Finally, in the very inner disc regions ($\lesssim 0.1$--$1\au$), temperatures are high enough to sublimate even the most refractory grains \citep{Muzerolle/etal/2003,DAlessio/etal/2004}.

%%%%%%%%%%%%%%%%
\section{Discussion}
\label{sec:discussion}
%%%%%%%%%%%%%%%%

Isotopic heterogeneity in meteorites and planetary materials may reflect both inheritance from the presolar molecular cloud and modification by a range of processes acting throughout the disc’s lifetime. Here, we review how each evolutionary phase—from cloud collapse to planetesimal formation—can imprint, preserve, or erase isotopic signals.

%%%%%%%%%%%%%%%%
\subsection{Inheritance from the molecular cloud}
\label{sec:inheritance}
%%%%%%%%%%%%%%%%

Isotopic heterogeneities observed in meteoritic samples may reflect inherited heterogeneity from the Sun’s nascent molecular cloud \citep{Clayton/1982,Dauphas/Marty/Reisberg/2002a,Dauphas/etal/2004,Nanne/etal/2019,Ek/etal/2020}. This heterogeneity could arise from variable enrichment by nearby stellar sources \citep{Lichtenberg/Parker/Meyer/2016}, such as supernovae and Wolf-Rayet stars, that polluted the molecular cloud prior to and during the Sun’s formation. Simulations suggest that between 7 and 30\% of Sun-like stars are contaminated by a single nearby supernova \citep{Pan/etal/2012}, and clumpy ejecta may introduce localised isotopic variations. Nonetheless, diffusion driven by hydromagnetic turbulence can homogenize gas over scales exceeding $0.3 \pc$ within $1 \Myr$ \citep{Pan/etal/2012}, and additional mixing occurs during the accretion process itself \citep{Kuffmeier/etal/2016}, raising the question of whether late-accreted material would differ significantly from earlier-accreted gas and dust, and whether any such differences could be preserved in the disc.

Dust grains are estimated to reside in the ISM for several hundred million years \citep{Gail/etal/2009}. During this time, they are subject to differential processing and destruction by interstellar shocks, grain-grain collisions, and sputtering, with the effects depending on grain size and composition \citep[e.g.][]{Hirashita/Yan/2009}. Numerical simulations of supersonic, dusty turbulence in molecular clouds show that grains in the sub-micron to micron size range undergo aerodynamic size-sorting, with larger micron-sized grains preferentially concentrating in regions of higher gas density \citep{Tricco/Price/Laibe/2017}. Presolar grains span a similar size range \citep{Liu/2025} and have comparable material densities \citep{Barthelmy/2018} to the silicate grains used in the simulations by \citet{Tricco/Price/Laibe/2017}, suggesting they would have experienced similar size-sorting during the molecular cloud phase. This is supported by \citet{Hutchison/etal/2022}, who observed distinct spatial distributions of presolar oxide and silicon carbide grains in the outer regions of their disc simulations, where gas densities are comparable to those in molecular clouds. If such sorting occurred before or during collapse, the spatial segregation of presolar grain types -- each carrying distinct isotopic compositions -- could produce a time-varying infall of isotopic signatures as material accretes onto the disc.

Infall onto protoplanetary discs has been observed across multiple stages of protostellar evolution -- from Class 0 \citep[e.g.][]{Thieme/etal/2022} to Class I \citep[e.g.][]{Cacciapuoti/etal/2024} and even Class II systems \citep[e.g.][]{Gupta/etal/2023}. In many systems, late-stage infall may contribute more than 50\% of the final disc mass \citep{Winter/Benisty/Andrews/2024}. Recent simulations indicate that infall rates sustained over Myr timescales can rival stellar accretion, effectively rejuvenating Class II discs and diminishing the importance of initially inherited properties \citep{Padoan/etal/2025}. These long-lived streamers connected to discs deliver material in a time-variable, non-axisymmetric manner and may carry dust that has bypassed dense envelope regions where substantial grain growth can occur \citep{Pineda/etal/2023}. Such delivery could modify the local size distribution and isotopic composition of grains within the disc. In the case of the Sun, \citet{Desch/Miret-Roig/2024} note that the protostellar disc might have accreted gas for several Myr, but conclude that conditions in the Sun’s natal environment, which likely limited the duration of gas accretion, make a shorter timescale of a few $\times 10^5 \yr$ more plausible. Regardless of the precise duration, the accretion phase appears to have been sufficiently long or variable to imprint spatial and/or temporal heterogeneities in the inherited presolar grain population \citep[e.g.][]{Jacquet/etal/2019,Nanne/etal/2019,Haba/etal/2021}. The final isotopic signature may thus reflect a competition between in-situ turbulent diffusion, which homogenises gradients, and episodic late infall, which can reintroduce heterogeneities (or vice versa with disc-driven heterogeneities being partially erased by infall).

Even under spatially homogeneous infall, thermal processing during accretion can fractionate solids by preferentially destroying thermally labile and isotopically anomalous carriers. This enriches the surviving material in more refractory components, imprinting isotopic heterogeneity onto initially well-mixed material -- either before or during infall \citep{Schiller/Paton/Bizzarro/2015,Van-Kooten/etal/2016}, or later after incorporation into the disc \citep{Trinquier/etal/2009,Akram/etal/2015}. However, this processing can be incomplete -- particularly in the outer disc -- allowing thermally labile carriers, including organics, to survive and retain a chemical and isotopic memory of the pre-disc environment \citep[e.g.][]{Chick/Cassen/1997,Drozdovskaya/etal/2016}. Alternatively, \citet{Ek/etal/2020} proposed that isotopic heterogeneities could arise from the preferential destruction of chemically labile mantles that condensed onto stardust in the ISM. In this scenario, the loss of isotopically homogenised mantle material would enhance the relative isotopic signal of the surviving, anomalous presolar cores.
In either case, thermal processing alters grain size and density, introducing radial variations in aerodynamic properties that influence the subsequent dynamical evolution of solids. 
In turbulent discs, remnants or refractory cores left behind after sublimation of volatile mantles in the hot inner regions are more likely to remain well mixed with the gas. In contrast, larger grains that retain more of their original mantles in the cooler outer disc can settle and drift inward, locally enhancing the dust-to-gas ratio.
Crucially, transient pressure enhancements driven by episodic infall could act as dust traps -- particularly when infall is massive and localised, the disc is low-mass and low-viscosity, and grains are resilient to fragmentation \citep{Zhao/etal/2025}. While the generic formation of planetesimals in pressure bumps has been widely studied \citep[e.g.][]{Drazkowska/Windmark/Dullemond/2013,Chambers/2021}, their potential connection to envelope dynamics offers a new unexplored link between infall history and early isotopic structuring in the Solar System.

%%%%%%%%%%%%%%%%
\subsection{Homogenising disc effects}
\label{sec:homogenisation}
%%%%%%%%%%%%%%%%

Despite the many mechanisms by which isotopic heterogeneities may be inherited or generated in the early disc, the subsequent viscous evolution during the $\sim1$--$5\Myr$ protoplanetary disc phase poses a significant barrier to their long-term preservation. Viscosity -- though still poorly understood in origin -- remains one of the dominant drivers of mixing. Observational constraints on disc accretion rates suggest that angular momentum transport must occur, likely driven by the magnetorotational instability \citep{Balbus/Hawley/1991}, gravitoturbulence \citep{Gammie/2001}, or hydrodynamic instabilities \citep{Lyra/Umurhan/2019}. In massive young Class 0 and I discs \citep[potentially more common than previously assumed][]{Schib/etal/2021}, gravitoturbulence can trigger a dynamo that amplifies and sustains magnetic fields \citep{Riols/Latter/2019}, even in the presence of strong non-ideal magnetohydrodynamic effects such as ohmic resistivity \citep{Deng/Mayer/Latter/2020} and ambipolar diffusion \citep{Riols/etal/2021}. In such discs, turbulence-driven diffusion is the dominant transport process \citep{Oosterloo/etal/2023} and has been shown to homogenise spatial heterogeneities of passive isotopic tracers down to levels of $\sim10\%$ \citep{Boss/2008}. Diffusion timescales can range from $\sim$100 years in the inner disc to $\sim10^5$ years in the outer disc and can be further accelerated by planet migration and radial surface flows \citep{Hutchison/etal/2022}. Supersonic surface accretion flows have been detected observationally \citep{Najita/etal/2021} and can be driven by magnetohydrodynamical (MHD) winds that extract angular momentum from the disc surface \citep[e.g.][]{Bai/2013,Okuzumi/2025}. These flows (see \cref{fig:vertical_cross-section}) can entrain small dust grains (whether primordial or fragments from larger grains) and rapidly redeposit them throughout the inner disc. Complete homogenisation of the inner disc can occur within $10^5$ years -- yielding mid-plane dust mineralogies that are far more radially mixed than predicted from equilibrium condensation models \citep{Oosterloo/Kamp/vanWestrenen/2025}.

By contrast, the later stages of disc evolution are increasingly thought to be characterised by lower levels of turbulence. Turbulence in discs is often conveniently characterised by a single dimensionless parameter, $\alpha_{\rm SS}$, called the Shakura-Sunyaev parameter \citep{Shakura/Sunyaev/1973}. This parameter expresses how effectively turbulence redistributes angular momentum, which in turn controls the overall rate of gas accretion through the disc. Recent surveys of Class II discs suggest that turbulence in these older systems is weaker than previously assumed, with typical values of $\alpha_{\rm SS}$ around a few $10^{-4}$ to $10^{-3}$ \citep{Rosotti/2023}. Dust evolution models support this picture: if turbulence is too strong ($\alpha \gtrsim 10^{-2}$), dust cannot remain trapped in observed ring-like structures \citep{de-Juan-Ovelar/etal/2016,Zormpas/etal/2022}, whereas if turbulence is too weak ($\alpha \lesssim 10^{-4}$), dust grains grow and concentrate too efficiently, producing rings in scattered light that are brighter than typically observed \citep{Rosotti/2023}. Even in relatively turbulent discs, magnetically dead zones in the inner disc \citep{Gammie/1996} can still experience reduced turbulence that suppresses mixing and accretion. For isotopic carriers, lower turbulence in discs has several important consequences. First, reduced stirring causes grains to (i) settle vertically more efficiently, (ii) drift radially more slowly, and (iii) collide at lower velocities, all of which promote growth into aggregates that preserve isotopic anomalies and are less susceptible to diffusion. Second, sharp radial contrasts in turbulence -- such as those at dead-zone edges -- create favourable conditions for planetesimal formation \citep{Drazkowska/Windmark/Dullemond/2013}, further sequestering isotopic signatures into larger bodies. Finally, as discs evolve and turbulence levels decline overall, late-stage accretion by streamers is subject to less diffusive mixing and homogenisation, making local isotopic signatures more difficult to erase.

The level of viscosity is also critical for planetary dynamics. Planets interact gravitationally with nearby gas, clearing material from their orbits and driving migration through torques from co-rotating material and viscous inflows \citep[e.g.][]{Lubow/Ida/2010}. One well-known example is the so-called Grand Tack model \citep{Walsh/etal/2011}, which invokes relatively high viscosities to reverse Jupiter’s inward migration upon capture into a 3:2 resonance with Saturn, thereby producing a small Mars and a compositionally diverse asteroid belt. Since then, however, numerous alternative models have been proposed to account for these features \citep[e.g.][]{Raymond/Izidoro/2017,Clement/etal/2018,Morbidelli/etal/2022}, and Solar System–like architectures have likewise been shown to arise in low-viscosity discs \citep{Griveaud/etal/2024}. Migration histories of giant planets are important because they shape the distribution of planetesimals in the disc -- scattering some into new orbits, shepherding others inward, or ejecting them entirely. Among other scenarios, gaps opened by planets further regulate radial transport and have been invoked to explain the isotopic dichotomy between carbonaceous and non-carbonaceous meteorites: Jupiter’s early formation may have acted as a long-lived barrier separating inner and outer reservoirs, preserving inherited heterogeneities over million-year timescales \citep{Kruijer/etal/2017,Alibert/etal/2018,Homma/etal/2024}. In low-viscosity discs, gaps are expected to be deeper and wider \citep{Ginzburg/Sari/2018}, potentially enhancing this separation of reservoirs -- but the outcome depends strongly on the migration behaviour. \citet{Lega/etal/2021} identified two migration modes in low-viscosity settings: (i) slow, stochastic migration driven by gap-edge vortices, and (ii) eccentricity growth in deep gaps, eventually allowing gas to leak through and reconnect previously isolated reservoirs. Moreover, small particles produced through mid-plane fragmentation and turbulent diffusion (both of which are influenced by viscosity) can still cross planetary gaps, providing a limited but non-negligible pathway for exchange \citep{Stammler/etal/2023}.

%%%%%%%%%%%%%%%%
\subsection{Thermal Processing Within the Disc}
\label{sec:thermal_processing_within_the_disc}
%%%%%%%%%%%%%%%%

\begin{figure}
    \centering{\includegraphics[width=\columnwidth]{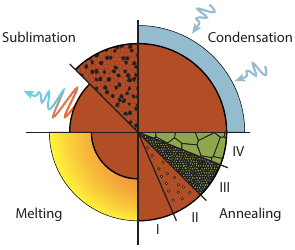}}
    \caption{Four principal modes of thermal processing affecting dust grains in protoplanetary discs. Solid brown regions represent an initial amorphous silicate grain prior to processing. Each quadrant illustrates potential outcomes as the grain experiences different temperature and pressure conditions. In colder regions of the disc, gases condense onto grain surfaces as ices. As a grain drifts inward toward the central star or undergoes a transient heating event, these ices and other volatile components sublimate, reducing the grain size or producing a more porous structure (represented by darkened holes). More refractory grains may either melt (partially or completely) and subsequently cool to form new crystalline phases, or they may undergo annealing. In the latter case, as temperature increases, seeds of new crystals (green regions) nucleate and grow within the amorphous material. These processes have important dynamical, collisional, and chemical consequences for individual grains and, when applied to large dust populations in the disc, can influence the isotopic signatures imprinted on Solar System bodies during the protoplanetary disc phase.
    }
    \label{fig:thermal_processing}
\end{figure}

Once material enters the protoplanetary disc, it is exposed to a range of thermal environments that can reprocess solids through evaporation, condensation, melting, and annealing (see \cref{fig:thermal_processing}). Given this broad evidence of thermally processed material in chondrites (see \cref{sec:constraints_from_meteorites}), it is not surprising that thermal processing in the discs has been evoked as a source for the systematic nucleosynthetic isotope variations in bulk meteorites illustrated in \cref{fig:rufenacht}. This idea has been presented in a variety of nuances. Initially, these nucleosynthetic patterns were interpreted as signatures of the thermal destruction of isotopically anomalous carriers, producing a residual heterogeneity visible on a bulk rock level in meteorites \citep{Trinquier/etal/2009}. For example, volatile Mo-bearing phases appear to have been preferentially destroyed, while refractory carriers of elements like W remained isotopically homogeneous \citep{Burkhardt/etal/2012,Burkhardt/Schonbachler/2015}. Similarly, correlations among isotopes hosted in different presolar phases (such as $^{84}$Sr and $^{54}$Cr) support selective thermal processing as the dominant mechanism \citep{Paton/Schiller/Bizzarro/2013}. Spatial gradients in thermal processing were also inferred from iron meteorites, which exhibit systematic Mo isotope differences linked to their formation distances from the Sun, consistent with heliocentric zoning of thermal conditions in the early disc \citep{Poole/Rehkaemper/Coles/2017}. Zirconium isotope variations suggest that silicate-based carriers of non-\textit{s}-process material were removed in the inner disc through similar heating \citep{Akram/etal/2015}. These data collectively imply that isotopic variations in meteorites reflect not a single homogenization or injection event, but complex, multi-stage thermal reprocessing.

A challenge/weakness of the thermal-processing model is that different carrier phases respond differently to heating, yet isotopic variations in unrelated elements are nevertheless well correlated in bulk meteorites (\cref{fig:rufenacht}). For instance, Ti isotope variations originating from distinct stellar production sites \citep{Hughes/etal/2008} are carried by mineralogically distinct phases in chondrites \citep{Trinquier/etal/2009} with different thermal stabilities. Yet, at the bulk-rock scale, these Ti isotope ratios are correlated (\cref{fig:rufenacht}). Moreover, these correlations extend to isotopes of other elements such as Cr, Ca and Ni, where isotopic anomalies occur in an even broader range of carrier phases -- such as nm-sized presolar spinel for Cr \citep{Nittler/etal/2018} and SiC for Sr, Zr, and Mo \citep{Dauphas/Marty/Reisberg/2002b,Schonbachler/etal/2005,Paton/Schiller/Bizzarro/2013} -- all with unique thermal responses and differing contributions to bulk variations.

To address this challenge, \citet{Ek/etal/2020} proposed an alternative to the destruction of isotopically anomalous carrier material: the preferential removal of abundant, isotopically solar-like material. Estimates suggest that $\sim97\%$ of the initial Solar System dust was reprocessed in the ISM, yielding average solar-system compositions, while the remaining $\sim3\%$ consists of anomalous presolar grains (SiC, nanodiamonds, graphite, silicates, oxides, etc.) \citep{Hoppe/Leitner/Kodolanyi/2017}. One possibility is that labile icy mantles that formed around refractory grains in the ISM \citep{Allamandola/etal/1999} were preferentially evaporated either during infall (\cref{sec:inheritance}) or in the hotter disc closer to the Sun \citep{Ek/etal/2020}. Removing solar-system solids increases the relative abundance of isotopically anomalous presolar grains. Importantly, the resulting isotopic signal scales with the amount of solar-system material removed and does not depend on the thermal stability of the anomalous phases, provided that temperatures remained below the threshold at which presolar grains are affected. This mechanism therefore naturally accounts for the correlations observed among isotope ratios within a single element and between different elements (\cref{fig:rufenacht}). The model is particularly effective in reproducing the correlated \textit{s}-process enrichments preserved in Mo, Zr, Ru, and Pd isotopes \citep{Ek/etal/2020}, for which Earth -- the innermost planet with available nucleosynthetic isotope compositions -- exhibits the strongest enrichments. However, this model alone cannot explain the depletion of supernova-derived isotopes (e.g. $^{48}$Ca, $^{50}$Ti, $^{54}$Cr) in the inner relative to the outer Solar System, which requires an additional process \citep{Van-Kooten/etal/2016,Ek/etal/2020}.

The idea of icy rim evaporation proposed by \citet{Ek/etal/2020} has been extended to models with repeated large-scale heating events, such as stellar outbursts \citep{Colmenares/etal/2024}. These models link early thermal episodes to evolving radial gradients in dust composition and isotopic signatures. In addition, local flash-heating events in the disc, such as those associated with chondrule formation, have been proposed to affect nucleosynthetic compositions \citep[e.g.][]{Frossard/etal/2021}. Taken together, these studies suggest that thermal processing can modify the isotopic composition of disc material, generating spatially variable and correlated signatures across the protoplanetary disc.

%%%%%%%%%%%%%%%%
\subsection{Dynamical Sorting Within the Disc }
\label{sec:dynamical_sorting}
%%%%%%%%%%%%%%%%

When thermal processing does alter the physical properties of grains (e.g. size, shape, composition, internal structure), it can also have a profound impact on their internal structure and, by extension, their collisional outcomes. For example, in chondritic material -- primarily composed of olivine, pyroxene, iron oxide, iron sulfide, metallic iron, and carbon-rich phases  \citep{Scott/Krot/2014}, although amorphous silicates are also present in less altered meteorites -- thermal evolution alters both composition and sticking behaviour. \citet{Bogdan/etal/2020} found that the surface energy of wet grains decreases steadily with temperature up to $\sim1250\K$, due to mineralogical changes such as the loss of metallic Fe, evolution of iron oxides, and redistribution of Fe into silicates. Around $\sim1300$--$1400\K$, they observed a sudden increase in constituent grain size that weakens aggregates, preventing further collisional growth and potentially promoting fragmentation. For dry grains, the sticking efficiency also decreases with temperature up to $\sim1000\K$ but is 10 times more efficient than wet grains \citep{Steinpilz/etal/2019}. Then, between $\sim1000$--$1400\K$, the trend reverses and increases exponentially by two orders of magnitude \citep{Pillich/etal/2021} (see illustration in \cref{fig:overview_macroscopic}). This confirms previous findings that water-ice-coated grains have lower sticking forces than previously assumed \citep{Gundlach/etal/2018}, and that the surface energy of amorphous H$_2$O ice is both lower than that of amorphous silicates \citep{Kimura/etal/2015} and temperature-dependent \citep{Gartner/etal/2017,Musiolik/Wurm/2019}. Moreover, it is not just the grain composition that matters: the atmosphere in which meteorite dust particles are heated also strongly influences compositional changes. For instance, experiments show that Fe in silicates is reduced to metallic Fe(Ni) in a hydrogen-rich gas at temperatures between $\sim1000$–$1200\K$ \citep{Pillich/etal/2023}, potentially enabling magnetic aggregation of ferromagnetic grains beyond the Curie line ($1041\K$) within the disc’s magnetic field, leading to the formation of Fe-rich planetesimals \citep{Kruss/Wurm/2018,Kruss/Wurm/2020,Bogdan/etal/2023}. Note, however, that since most primitive silicates (e.g. forsterite and enstatite) are Fe-poor, this mechanism would likely only become effective after some dust processing had occurred.

The physical changes induced by thermal processing in grains have implications not only for their collisional behaviour but also for their aerodynamic properties. For example, the subsequent dust sorting of altered and unaltered grains can be an important mechanism for generating nucleosynthetic variations \citep[e.g.,][]{Steele/etal/2012,Schonbachler/etal/2025}. Studies of chondrule sizes suggest that aerodynamic sorting led to similar size ranges of chondrules and refractory inclusions within each chondrite group \citep{Cuzzi/etal/2001,Jones/2012}. Most presolar grains with their extreme nucleosynthetic compositions are $\nm$- to $\gtrsim \mum$-sized. In contrast, larger particles -- such as single minerals, chondrules, and refractory inclusions -- generally exhibit more solar-system–like isotopic compositions (although refractory inclusions show stronger deviations from solar). Based on size alone, these two populations of dust grains would encounter very different dynamics within the disc (see \cref{sec:dynamical_evolution}).

Building on this distinction, \citet{Hutchison/etal/2022} demonstrated that nucleosynthetic variations can arise naturally in discs -- even if they begin from a homogeneously mixed state -- provided two or more dynamically distinct grain populations differ isotopically. Applied to dust in chondrites, this implies that the fine-grained dust containing presolar grains (as preserved, for example, in CI chondrites) was homogeneously distributed in the disc, while varying fractions of more solar-system–like material were locally concentrated in different regions through interactions with the gas. This variable dilution of the presolar grain fraction enhances or reduces the isotopic signal and inherently leads to correlated isotopic heterogeneities. In the model of \citet{Hutchison/etal/2022}, grain growth and fragmentation were neglected, which would affect the magnitude of the variations by mixing anomalous and solar-like material in the different dust populations. However, within this framework, they predicted that the outer disc (including cometary material) is preferentially enriched in presolar grains (including comets) -- a prediction supported by evidence from interplanetary dust particles (IDPs) and specific clasts in returned samples from asteroid Ryugu \citep{Barosch/etal/2022,Nguyen/etal/2023}. The central mechanism responsible for this outcome in their simulations was size sorting of grains driven by viscous evolution and backreaction, processes that operate even in smooth discs. Large grains migrate radially inward under gas drag, while their backreaction on the gas, together with viscous expansion, carries small, tightly coupled grains outward. Dust traps such as rings, spirals, and vortices further enhance size sorting in discs, preferentially concentrating large grains while leaving small grains more diffusely distributed. Observations support this picture of differential sorting: scattered-light imaging with SPHERE, GPI, and HiCIAO reveals extended distributions of small grains \citep[e.g.][]{Garufi/etal/2017,Benisty/etal/2023}, while (sub-)millimetre observations with ALMA show thinner, sharper substructures traced by large grains \citep[e.g.][]{Andrews/etal/2018,Dullemond/etal/2018}. These substructures (i.e. dust traps) provide natural sites for planetesimal formation via the streaming instability \citep[e.g.][]{Lau/etal/2022}, effectively locking local isotopic signatures into larger bodies.

Dynamically, planetesimals differ starkly from small grains: they are largely decoupled from the gas, with drag acting mainly to damp eccentricity and inclination over long timescales (as opposed to driving diffusive mixing and migration). Their evolution is instead governed by mutual gravitational interactions \citep[e.g. viscous stirring, dynamical friction, and collisions; see][]{Kokubo/Ida/2012} and potentially surface erosion from gas \citep{Schaffer/etal/2020}. Although chondrites preserve an invaluable record of early Solar System material, most show evidence of brecciation \citep{Bischoff/etal/2006} and aqueous alteration or metasomatism/metamorphism \citep{Brearley/2003,Huss/etal/2003}. Collisions among first generation planetesimals -- especially in the later stages of the disc -- may have liberated fragments of primitive material back into the dust population that could later be incorporated, along with second generation dust, in the formation of new planetesimals. Evidence of such incorporation is recoded in chondrites, but only represents a minor component \citep[e.g.][]{Bischoff/etal/2006}. Moreover, the tight clustering of isotopic signatures across meteorite groups suggests that large-scale mixing by collisions was limited; however, this may have contributed to the subtle spread observed within individual clusters.

%%%%%%%%%%%%%%%%
\section{Summary}
\label{sec:summary}
%%%%%%%%%%%%%%%%

Despite significant progress in understanding individual processes such as infall, thermal processing, and grain sorting, their combined impact on nucleosynthetic variations is only beginning to be explored. Untangling how these processes interact (see \cref{fig:overview_macroscopic}) to shape the isotopic architecture of the Solar System remains a key challenge for future research. While the precise origin of nucleosynthetic heterogeneities remains uncertain, we emphasise that they were almost certainly modified by mechanisms operating within the disc. These mechanisms can be broadly classified into three categories: (i) dynamical, (ii) collisional, and (iii) thermal. In our discussion we have focussed mainly on the first and third categories, as they are somewhat easier to constrain observationally. Thermal processing through accretion, shocks, radial and vertical temperature gradients, or stellar outbursts has been considered in previous studies as possible contributors/modifiers of isotopic heterogeneities, though often in isolation from other disc processes. In contrast, dynamical size-sorting via turbulent diffusion, vertical settling, viscously modified radial migration, backreaction, and dust trapping has received comparatively little attention, despite being a robust and unavoidable feature of disc evolution. This is all the more important given the recent results of \citet{Hutchison/etal/2022}, which revealed that isotopic diversity can emerge naturally through size sorting, even without any inherited heterogeneity. In this light, the disc may not have been the passive, homogenising reservoir it was once thought to be, but rather an active engine of isotopic evolution, reinforcing and even generating new fractionation with time. Nevertheless, disc processing alone is unlikely to explain the full spectrum of observed heterogeneities. Contamination of the parental molecular cloud with newly generated isotopes from massive stars (e.g. Wolf–Rayet stars) and core-collapse supernovae remains a possibility, as evidenced by the presence of short-lived radionuclides such as $^{26}$Al and $^{53}$Mn \citep[e.g.][]{Haba/etal/2021,Desch/etal/2023a}. This material would have been incorporated into the disc during infall or injected later via streamers, suggesting that disc processing does not erase all of its inherited heterogeneities. Understanding the isotopic architecture of the Solar System may therefore require accounting for contributions and modifications across the full evolutionary arc of the disc, from initial infall to its final dispersal.

In this review we have focused on the protoplanetary disc phase; however, the long subsequent dynamical evolution during the debris-disc phase could have led to additional mixing. Following gas dissipation, energetic collisions -- evident from the meteorite record \citep{Hunt/etal/2022} -- generated both dust and larger fragments, some of which were incorporated into second-generation asteroids and pre-existing planetary bodies. In this later stage, dynamical evolution was shaped not only by gravitational perturbations and interactions with large planets, but also by size-dependent forces acting on smaller solids. Sub-$\mum$ to $\cm$-sized grains were removed from the Solar System by radiation pressure, Poynting–Robertson drag, and/or solar-wind (corpuscular) drag \citep{Burns/Lamy/Soter/1979}, while the spins and orbits of boulders and asteroids were influenced by the Yarkovsky and YORP effects \citep{Paddack/1969,Rubincam/2000}. These processes were particularly important for rubble-pile asteroids such as 2008 TC$_3$ \citep[the progenitor of the Almahata Sitta strewn field in the Nubian desert;][]{Bischoff/etal/2022}, as well as for Bennu and Ryugu, which have been visited by recent sample-return space missions \citep{Tachibana/etal/2022,Lauretta/etal/2024}.

%%%%%%%%%%%%%%%%%%%%%%%%%%%%%%%%%%%%%%%%%%%%%%%%%%


%%%%%%%%%%%%%%%%%%%%%%%% referenc.tex %%%%%%%%%%%%%%%%%%%%%%%%%%%%%%
% sample references
% %
% Use this file as a template for your own input.
%
%%%%%%%%%%%%%%%%%%%%%%%% Springer-Verlag %%%%%%%%%%%%%%%%%%%%%%%%%%
%
% BibTeX users please use
\bibliographystyle{mnras}
\bibliography{bibliography}
\end{document}